%Paper: hep-th/9406159
%From: kelley@miu.edu (Steve Kelley)
%Date: Thu, 23 Jun 1994 20:37:01 +0600

\font\titlefont=cmbx12
at 14pt \vsize=52truepc\voffset=1pc \hsize=37truepc\hoffset=1pc \pageno=1
\footline{\ifnum\pageno>1\hss\lower.25in\hbox{\tenrm\folio}\hss\fi}

 \widowpenalty=1000 \advance\baselineskip by 2pt

\bgroup \advance\baselineskip by 2pt \null\vskip-2\baselineskip \rightline{\it
}
\vskip\baselineskip
\rightline{ \it To appear in the Journal of Mathematical Physics }
\bigskip
\rightline{ MIU-THP-67/94}
 \rightline{ May, 1994}
\vskip.5in
\centerline{\titlefont On Transformations Preserving the Basis}
\centerline {\titlefont Conditions of a Spin Structure Group in}
\centerline {\titlefont Four-Dimensional Super String Theory in}
\centerline{\titlefont Free Fermionic Formulation}
\vskip 0.3in
\centerline {Valery A. Kholodnyi}
\vskip 0.2in
\centerline {Department of Physics}
\centerline { Maharishi International University}
\centerline {Fairfield, Iowa 52557}

\vskip.4in \egroup

\body \parindent=25pt \parskip\medskipamount \advance\baselineskip by 6pt

\centerline {\bf Abstract}
\vskip 0.1in
\bgroup
\leftskip 0.7in\rightskip=\leftskip
\noindent Let  $\Xi$  stand for a finite abelian spin structure group of
four-dimensional superstring theory in free fermionic formulation whose
elements are
64-dimensional  vectors (spin structure vectors) with rational entries
belonging to
$\rbrack -1,\, 1\rbrack $ and the group operation is the $mod\, \, 2 $ entry by
entry
summation $\oplus $ of these vectors. Let $B=\{b_i,\, i= 1,\cdots ,k+1\}$ be a
set of
spin structure vectors such that  $b_i$ have only entries 0 and 1 for any $\,
i=
1,\cdots ,k$, while $b_{k+1}$ is allowed to have any rational entries belonging
to
$\rbrack -1,\, 1\rbrack $ with even $N_{k+1}$, where $N_{k+1}$ stands for the
least
positive integer such that $N_{k+1}b_{k+1}= 0\,mod\,2$. Let $B$ be a basis of
$\Xi$,
i.e., let $B$ generate $\Xi$, and let
$\Lambda_{m, n}$ stand for the transformation of $B$ which replaces $b_n$ by
$b_m\oplus
b_n$  for any $m \ne k+1$, $n \ne 1$, $m \ne n$. We prove that if $B$ satisfies
the
axioms for a basis of spin structure group $\Xi$, then $B'=\Lambda_{m, n}B$
also
satisfies the axioms. Since the transformations $\Lambda_{m,n}$ for different
$m$ and
$n$ generate all nondegenerate transformations of the basis $B$ that preserve
the vector
$b_1$ and a single vector $ b_{k+1} $ with general rational entries, we
conclude that
the axioms are conditions for the whole group $\Xi$ and not just conditions for
a
particular choice of its basis. Hence, these transformations generate the
discrete
symmetry group of four-dimensional superstring models in free fermionic
formulation.
This is the physical meaning of the obtained result. The practical impact is
that in
searching for a realistic model in four-dimensional superstring theory in free
fermionic
formulation it is enough to search only through different groups $\Xi$, which
number
dramatically less than the number of their different bases.\par
\egroup

\bigskip
\beginsec 1. Introduction

In searching for a realistic model in four-dimensional superstring theory in
free
fermionic formulation [1 and 2], one faces the following problem: despite the
fact
that the GSO projection does not depend on a particular choice of a basis for a
finite
abelian spin structure group $ \Xi $, the conditions for the group itself
heavily depend
on this choice. These conditions are formulated in terms of six axioms for the
basis
which we will present later for the convenience of reference. Thus it might
seem that
the whole physical theory depends not only on the group $ \Xi $ but also on a
particular choice of its basis. This situation is very unsatisfactory from both
pure
theoretical perspective and from practical stand point of searching for a
realistic
four-dimensional superstring model in free fermionic formulation where one has
to search
not only through different groups $\Xi $ but also through their different bases
which
number is nearly ``countless".

In the present paper we show that this is not the case: the conditions for the
basis of
the group $\Xi $ are invariant with respect to some natural transformations of
the basis
and hence are, in fact, conditions for the whole group $\Xi $. In order to
formulate our
main result let us, following references [1 - 4], recall necessary facts and
terminology from four-dimensional superstring theory in free fermionic
formulation.

The elements of the finite abelian spin structure group $\Xi \cong
Z_{N_1}\times
Z_{N_2}\times \cdots \times Z_{N_k}$ are referred to as spin structure vectors
and are
represented as 64-dimensional vectors with rational entries belonging to
$\rbrack -1,\,
\, 1\rbrack $. The group operation is the $mod\, 2$ entry by entry summation of
these
vectors which we hereafter denote as $\oplus $. The group $\Xi $ is generated
by a
basis, i.e. a set of spin structure vectors $B=\{b_i, \, i= 1,\cdots , k\}$ in
such a way that $N_i$ is the least positive integer such that $N_ib_i=0\,mod\,
2$ and
any  spin structure vector $b\in \Xi $ can be represented as:
$$b=a_1b_1\oplus a_2b_2\oplus\cdots\oplus a_kb_k,$$
where $a_i= 0,\cdots ,N_i-1$, and hereafter, if it is not ambiguous, a
multiplication by an integer is understood in $mod\, 2$ sense.

The $l$-th entry $b^l$ of any spin structure vector $b\in \Xi $ corresponds to
the so-called boundary condition of the $l$-th fermion. Integer entries of a
spin
structure vector, i.e., entries equal to 0 and 1 correspond respectively to the
so-called antiperiodic and periodic fermions.

{\bf Remark:} The above terminology comes from the following physical picture
[1, 2]:
in four-dimensional string theory in free fermionic formulation there are
considered 64
fermions. They reside on a so-called world sheet which, in the $g$-th term in
perturbation expansion, is a genus $g$ Rieman surface with $2g$ noncontractable
loops.
The parallel transport of the $l$-th fermion around such loops change it by a
phase
factor $-e^{i\pi b^l}$. Thus a spin structure vector $b$ with entries $b^l,\,
l= 1,\cdots ,64$ serves to specify boundary conditions for the set of 64
fermions
with respect to such parallel transports.

Any spin structure vector as a 64-dimensional vector has the following
structure: the
first twenty entries correspond to the so-called left moving fermions and the
remaining
44 entries correspond to the so-called right moving fermions. Among the left
moving
fermions two first entries correspond to the so-called space-time fermions and
they can
be either both equal to 0 or to 1. In the first case, a spin structure vector
is said to
belong to the Neveu-Schwarz sector while in the second case to the Ramond
sector. The
remaining 18 entries that correspond to the left moving sector are formed by 6
triplets
$(\chi _i, y_i, \omega _i),\enskip i= 1,\cdots ,6$ in the following way:
$b^i=\chi
_i$, $ b^{6+i}=y_i$ and $b^{12+i}=\omega _i $.

There are two types of fermions: real and complex ones. For any spin structure
vector $b$ each real fermion is described by one entry $b^l$ which can be
either 0 or
1, and each complex fermion is described by a pair of entries $b^{l'}=b^{l''}$,
$ l'\ne
l''$, which can be any rational number from $\rbrack -1,1\rbrack$. Let us
comment that
for different spin structure vectors entries corresponding to the same real or
complex
fermion, i.e., entries corresponding to the same indices $l$ or $l',\, l''$
might be
different.

In the present paper according to the most successful four-dimensional
superstring
models in free fermionic formulation (see for example references [3], [5] and
[6]), we
assume the left moving fermions to be real, i.e., the first 20 entries of any
spin
structure vector to be either 0 or 1, while we assume all the entries that
correspond
to complex fermions to be among the right moving fermions, i.e. to be among the
last 44
entries.

{\bf Remark:} Since complex fermions are allowed to have pairs of entries both
equal to
0 or 1, in some string models  authors complexify pairs of real left-moving
fermions or
even one left-moving fermion with one right-moving fermion to form the
so-called Ising
model [7]. However, what is important for us is that any entry different from 0
and 1
can be found only among the last 44 entries.

We denote two special spin structure vectors with all entries equal to 0 and 1
by $\bf 0$
and $\bf 1$ respectively.

There is an inner product defined on the spin structure group $\Xi $:
$$(\alpha; \beta)={1\over
2}(\sum_{l=1}^{20}-\sum_{l=21}^{64})\alpha^l\beta^l,\qquad
\forall \, \alpha ,\beta \in \Xi.\eqno {(1)}$$
To make notation short we will write the above definition as:
$$(\alpha ;\beta )={1\over 2}\sum_l \alpha^l\beta^l.$$

Now we are ready to present the set of axioms for a basis B:

{\bf Axioms:}

\item{$A_1.$} The basis $B=\{b_i,\, i= 1,\cdots ,k\}$ has to be such that:
$$a_1 b_1\oplus a_2 b_2\oplus\cdots\oplus a_kb_k=0\iff a_i=0\, mod\, N_i,\,
\forall
i= 1,\cdots ,k;$$

\item{$A_2.$}There is such vector $b_1\in B$ that:
 $${1\over 2}N_1b_1={\bf 1}\, mod\, 2;$$

\item{$A_3.$}Let $N_{ij}$ be the least common multiple of $N_i$ and $N_j$,
then:
$$N_{ij}(b_i; b_j)=0\, mod\, 4,\qquad \forall i,j=1,\cdots ,k;$$

\item{$A_4.$} For any $i=1,\cdots ,k$:
  $$N_i(b_i; b_i) =
    \cases{0\, mod\, 4,& $N_i$  is  odd\cr
      				 0\, mod\, 8,& $N_i$ is  even\cr
           N_i({\bf 1};{\bf 1})\, mod\,16,
                                    & $N_i(b_i+{\bf 1})=0\, mod\, 4$;\cr }$$

\item{$A_5.$}The number of real fermions that are simultaneously periodic, i.e,
the
number of entries that are simultaneously equal to 1 in any three basis spin
structure vectors $b_i,\, b_j$ and $b_f$ is even;

\item{$A_6.$}Any given triplet $(\chi _i, y_i,\omega _i),\, i= 1,\cdots ,6$ in
any
basis spin structure vector contains an odd number of periodic fermions, i.e.,
an odd number of entries 1 if this vector belongs to the  Ramond sector and an
even
number of periodic fermions, i.e., an even number of entries 1 if this vector
belongs to
the Neveu-Schwarz sector.

\medskip

Further after we will assume, as it is usually done in physical literature (see
for
example reference [2]), that $b_1={\bf 1}$. It is obvious that {\bf 1}
satisfies
axiom $A_1$.

Let us introduce the following notation: we denote as $\Lambda_{m,n}$ a
transformation of a set of spin structure vectors $B=\{b_i,\, i= 1,\cdots ,k\}$
which
replaces the vector $b_n$ by the vector $b_m\oplus b_n$ and does not change the
rest of
the vectors.

{\bf Remark:} It is obvious, that $\Lambda_{m,n}$ for different $m,n= 1,\cdots
,k$
such that $m\ne n$ generate all nondegenerate transformations of the set B.

{\bf Definition 1:} We will say that the basis $B=\{b_i,\,i= 1,\cdots ,k\}$ of
the
group $\Xi $ is a proper basis  if it satisfies axioms $A_1-A_6$.

{\bf Definition 2:} A spin structure vector is called an integer one if it only
has
entries equal to 0 or 1. A spin structure vector that has at least one entry
not
equal to 0 or 1 is called a non-integer one.

Now we are ready to state the main result of this paper:

{\bf Theorem}: Let $B=\{b_i,\, i= 1,\cdots , k+1\}$ be a proper basis of a
finite
abelian spin structure group $\Xi$ such that $b_i$  are integer spin structure
vectors
for any $i=1,\cdots ,k$ with $b_1={\bf 1}$ and $b_{k+1}$ is a non-integer
vector.
Let $N_{k+1}$ be even. Then $B'=\Lambda_{m,n}B$ is also a proper basis of $\Xi
$ for any
$ m,n= 1,\cdots ,k+1$ such that $m\ne k+1,\, n\ne 1$ and $m\ne n$.

{\bf Remark:} If $N_{k+1}$ would be odd, then the axiom $A_1$ would not be
satisfied. The reason for this is that the following ``linear" combination with
non-zero coefficients $N_{k+1}(b_{k+1}\oplus b_m)\oplus b_m=0\, mod\,2$ for any
$m= 1,\cdots ,k$. Indeed, it is easy to see that for an odd $N_{k+1}$ the least
positive integer $N_{k+1}'$ such that $N_{k+1}'(b_{k+1}\oplus b_m)=0\, mod\, 2$
is equal
to $2N_{k+1}$, and clearly $N_{k+1}\ne 0\, mod\, N_{k+1}'$. At the same time,
by
definition of $N_{k+1}$, $N_{k+1}b_{k+1}=0\, mod\, 2$, and since $N_{k+1}$ is
odd,
$N_{k+1}b_m=b_m\, mod\, 2$ for any $m=1,\cdots ,k$. Thus,
$N_{k+1}(b_{k+1}\oplus b_m)\oplus b_m\mathop{=}\limits ^{mod\,2}\,b_m\oplus
b_m=0$.

Let us make a few comments on the Theorem. First we would like to point out
that the
most successful four-dimensional superstring models in free fermionic
formulation (see
for example  references [3],[5] and [6]) are built with bases $B$ which have
only one
non-integer vector. That is why the most physically interesting transformations
$\Lambda_{m,n}$ are those that leave the number of non-integer basic spin
structure
vectors equal to one, i.e, $m\ne k+1$. The condition $n\ne 1$ means that
according to
the axiom $A_2$, we are not allowed to change vector $b_1={\bf 1}$. Under this
restriction and taking into account the above Remark, the Theorem states that
the
axioms $A_1-A_6$ are conditions for the whole group $\Xi$ and not the
conditions for a
particular choice of its basis.

\bigskip
\centerline {\bf 2. Proof of the Theorem}

In order to prove the Theorem we need some preparations:

{\bf Lemma}: Let $B_I=\{ b_i,\, i= 1,\cdots ,k\}$ be a proper basis of a finite
abelian spin structure group $\Xi _I$ such that $b_i$ are integer spin
structure vectors
for any $i= 1,\cdots ,k$. Then $B'_I=\Lambda_{m,n}B$ is also a proper basis of
$\Xi
_I$ for any $m,n=1,\cdots ,k$ such that $ n\ne 1$ and $ m\ne n$.

{\bf Remark:} It is obvious, that the Lemma is a direct consequence of the
Theorem.
However, in addition to the fact that we need the result of the Lemma in order
to prove
the Theorem, the case considered in the Lemma is important in four-dimensional
superstring model building in free fermionic formulation by itself [8]. That is
also why
we present the Lemma as a separate result.

{\bf Proof of the Lemma}: Let us notice first that the set of spin
structure vectors $B'_I=\Lambda_{m,n}B$ for any $m,n= 1,\cdots ,k$ such
that $n\ne 1$ and $ m\ne n$ is a basis, i.e., generates the group $\Xi_I$.

Let $N'_n$ stand for the least positive integer such that $N'_nb'_n=0\, mod\,
2$ where
$b'_n=b_m\oplus b_n$. Since $b_m$ and $b_n$ are integer vectors, $b'_n$ is also
an
integer vector for any $m,n= 1,\cdots ,k$. It is easy to see that $N_i=N'_n=2$
for
any $i,n,m= 1,\cdots ,k$.

We are going to prove the Lemma axiom by axiom.

{\bf Axiom $A_1$}: Let us show directly that if the basis $B_I$ satisfies axiom
$A_1$
then the basis $B'_I$ also does. In order to do it, let us consider the
following
``linear" combination:
  $$a_1b_1\oplus\cdots\oplus a_nb'_n\oplus\cdots\oplus a_kb_k$$
where $a_i= 0,1$ for any $i= 1,\cdots ,k$. It is clear that if all
$a_i=0$ then the above ``linear"combination is also equal to 0. Let us
prove the converse statement. In order to do it, we rewrite the above``linear"
combination as follows:
  $$a_1b_1\oplus\cdots\oplus a_nb_n\oplus\cdots\oplus(a_m\oplus
  a_n)b_m\oplus\cdots\oplus a_kb_k.$$
Since by the assumption the basis $B_I$ satisfies axiom $A_1$ then from the
equality of
the above ``linear" combination to zero it follows that $a_i=a_m\oplus a_n=0$
for any
$i= 1,\cdots ,k$ such that $i\ne m$. Thus, only $a_m$ is left to be proved to
be
equal to zero. But, it directly follows from the equalities $a_n\oplus a_m=0$
and
$a_n$=0.

{\bf Axiom $A_2$:} Axiom $A_2$ obviously holds since the transformations
$\Lambda_{m,n}$
do not change the spin structure vector $b_1$ due to the condition $n\ne 1$.

{\bf Axiom $A_3$:} Let us notice first that the least common multiple of all
$N_i$ and
$N'_n$ is equal to 2. That is why what we have to prove is that:
  $$2(b_i; b_m\oplus b_n)=0\, mod\, 4,\qquad \forall \, i, m, n= 1,\cdots ,k.$$
In order to do this, it is enough to show that:
  $$2(b_i; b_m\oplus b_n)\mathop{=}\limits^{mod\, 4}\, 2(b_i; b_m+b_n),\qquad
  \forall \, i, m, n= 1,\cdots ,k.\eqno{(2)}$$
Indeed, due to the linearity of the inner product defined in (1) with respect
to the
regular summation, we rewrite the right-hand side of the above equality as
follows:
  $$2(b_i; b_m+ b_n)=2(b_i; b_m)+2(b_i; b_n).$$
The right-hand side of the above expression is equal to $0\, mod\,4$ for any
$i,m,n
= 1,\cdots ,k$ since each of the summands is equal to $0\, mod\,4$ because the
vectors $b_i, b_m$ and $b_n$ belong to the basis $B_I$ which is assumed to
satisfy
axioms $A_3$ and $A_4$.

Let us prove equality (2). In order to do it, we split the sum in the inner
product as
follows:
  $$2(b_i, b_m\oplus b_n)=\sum_l b_i^l(b_m^l\oplus
  b_n^l)=\mathop{\sum}\limits_{l:\, b_m^l=b_n^l}\, b_i^l(b_m^l\oplus b_n^l)+
  \mathop{\sum}\limits_{l:\, b_m^l\ne b_n^l}\, b_i^l(b_m^l\oplus b_n^l).$$
We consider the sums in the right-hand side of the above expression separately.
The
first sum is equal to zero, since the condition $b_m^l=b_n^l$ means that either
$b_m^l=b_n^l=0$ or $b_m^l=b_n^l=1$, and in both cases $b_m^l\oplus b_n^l=0$.
For the
second sum, it is easy to see that $b_m^l\oplus b_n^l=b_m^l+b_n^l$, since the
condition
$b_m^l\ne b_n^l$ means that either $b_m^l=1$ and $b_n^l=0$ or $b_m^l=0$ and
$b_n^l=1$.
Therefore, we obtain the following equality:
  $$2(b_i, b_m\oplus b_n)=\mathop{\sum}\limits_{l:\, b_m^l\ne b_n^l}\,
  b_i^l(b_m^l+ b_n^l).$$
In order to finish the proof, it is enough to notice that the right-hand side
of
the above relation differs from the right-hand side of equality (2) by the
following
expression:
  $$\mathop{\sum}\limits_{l:\, b_m^l=b_n^l=b_i^l=1}\, b_i^l(b_m^l+ b_n^l)$$
which is equal to $0\, mod\, 4$ for any $i,m,n= 1,\cdots ,k$. Indeed, each
summand in
the above sum is equal to 2 and the number of summands is even since it is
equal to the
number of common entries 1 in vector $b_i$, $b_m$ and $b_n$ which is even for
any
$i,m,n= 1,\cdots ,k$ due to axiom
$A_5$ .

{\bf Axiom $A_4$:} First, let us show that $N_n'(b_m\oplus b_n+{\bf 1})\ne 0\,
mod\,4$
for any  $m,n= 1,\cdots ,k$ so the condition $N_n'(b_m\oplus b_n)=N_n'({\bf
1};{\bf
1})\, mod\, 16$ is never applied. Indeed, since  $b_m$ and $b_n$ are integer
vectors
and $N_n'=2$ for any $n= 1,\cdots ,k$, the above equality holds if and only if
$b_m\oplus b_n={\bf 1}$. But, since $b_1={\bf 1}$, it would contradict axiom
$A_1$
because the following ``linear" combination with non-zero coefficients
$b_m\oplus
b_n\oplus b_1=0$.

Now let us prove that:
  $$2(b_m\oplus b_n; b_m\oplus b_n)=0\, mod\, 8,\qquad \forall m,n=1,\cdots
,k.$$
In order to do this, it is enough the show that:
  $$2(b_m\oplus b_n; b_m\oplus b_n)\mathop{=}\limits^{mod\, 8}\,
  2(b_m+b_n; b_m+b_n),\qquad \forall \, m,n= 1,\cdots ,k.\eqno{(3)}$$
Indeed, due to the linearity of the inner product defined in (1) with respect
to the
regular summation we rewrite the right-hand side of the above equality as
follows:
  $$2(b_m+b_n; b_m+b_n)=2(b_m; b_m)+4(b_m; b_n)+2(b_n; b_n).$$
The right-hand side of the above expression is equal to $0\, mod\, 8$ for any
$m,n=
1,\cdots ,k$. Indeed the first and the third summands are equal to $0\, mod\,
8$ for any
$m,n= 1,\cdots ,k$ because the vectors $b_m$ and $b_n$ belong to the basis
$B_I$
which is assumed to satisfy axiom $A_4$. The second summand is equal to
$0\,mod\, 8$
because $2(b_m;b_n)=0\, mod\, 4$ for any $m,n= 1,\cdots ,k$ due to axioms $A_3$
and
$A_4$.

Let us prove equality (3). In order to do it, we split the sum in the inner
product as
follows:
  $$\eqalignno{
  2(b_m\oplus b_n; b_m\oplus b_n)&=\sum_l(b_m^l\oplus b_n^l)(b_m^l\oplus
  b_n^l)\cr
  &=\mathop{\sum}\limits_{l:\, b_m^l=b_n^l}\, (b_m^l\oplus b_n^l)(b_m^l\oplus
  b_n^l)+\mathop{\sum}\limits_{l:\, b_m^l\ne b_n^l}\, (b_m^l\oplus
b_n^l)(b_m^l\oplus
  b_n^l).\cr}$$
We consider the sums in the right-hand side of the above equality separately.
The first
sum is equal to zero, since the condition $b_m^l=b_n^l$ means that either
$b_m^l=b_n^l=0$ or $b_m^l=b_n^l=1$, and in both cases $b_m^l\oplus b_n^l=0$.
For the
second sum, it is easy to see that $b_m^l\oplus b_n^l=b_m^l+b_n^l$ since the
condition
$b_m^l\ne b_n^l$ means that either $b_m^l=1$ and $b_n^l=0$ or $b_m^l=0$ and
$b_n^l=1$. Therefore, we obtain the following equality:
  $$2(b_m\oplus b_n; b_m\oplus b_n)=\mathop{\sum}
  \limits_{l:\, b_m^l\ne b_n^l}\, (b_m^l+b_n^l)(b_m^l+b_n^l).$$
In order to finish the proof, it is enough to notice that the right-hand side
of
the above relation differs from the right-hand side of equality (3) by the
following
expression:
  $$\mathop{\sum}\limits_{l:\, b_m^l=b_n^l=1}\, (b_m^l+b_n^l)(b_m^l+b_n^l).$$
which is equal to $0\, mod\, 8$ for any $m,n = 1,\cdots ,k$. Indeed, each
summand in
the above sum is equal to 4 and the number of the summands is even since it is
equal
to the number of common entries 1 in the vectors $b_m$ and $b_n$, which is even
for any
$m,n= 1,\cdots ,k$ due to axiom $A_5$.

{\bf Axiom $A_5$:} Let us consider an abstract set $F$ of 64 elements and the
set $2^F$ of subsets of F. Following [1] let us consider the isomorphism
$S:\Xi_I\to 2^F$ defined as follows: if the l-th entry of a spin structure
vector
$b\in\Xi_I$ is equal to 1 (0), then the element of $F$ belongs (does not
belong)
to a subset $S(b)$. It is easy to see that the group operation $\oplus$ in
$\Xi_I$
becomes, under the isomorphism $S$, the symmetric difference of sets:
  $$S(b\oplus b')=S(b)\bigtriangleup S(b'),\qquad \forall \, b, b'\in \Xi_I,$$
where $S(b)\bigtriangleup S(b')=S(b)\cup S(b')-S(b)\cap S(b')$ and the minus
sign
stands for the set difference.

In terms of the isomorphism S, what we need to show is that the spin structure
vector
$S^{-1}\circ \bigl( S(b_i)\cap S(b_j)\cap S(b_m\oplus b_n)\bigr) $ for any
$i,j, m, n= 1,\cdots ,k$ has an even number of entries 1 if the spin structure
vector $S^{-1}\circ \bigl( S(b_i)\cap S(b_j)\cap S(b_f)\bigr) $ has even number
of
entries 1 for any $i, j, f= 1,\cdots ,k$. Reexpressing $S(b)\bigtriangleup
S(b')$ as
$\bigl( S(b)-S(b)\cap S(b')\bigr) \cup \bigl( S(b')-S(b)\cap S(b')\bigr) $ and
using the
elementary properties of the operations $\cup $, $\cap$ and $-$ we obtain the
following
chain of equalities:
  $$\eqalignno{
  &S^{-1}\circ\lbrack S(b_i)\cap S(b_j)\cap S(b_m\oplus b_n)\rbrack= \cr
  &=S^{-1}\circ\lbrack S(b_i)\cap S(b_j)\cap \{ \bigl( (S(b_m)-S(b_m)\cap
  S(b_n)\bigr) \cup\bigl( S(b_n)-S(b_m)\cap S(b_n)\bigr) \} \rbrack \cr
  &=S^{-1}\circ\lbrack \{ S(b_i)\cap S(b_j)\cap S(b_m)-S(b_i)\cap
  S(b_j)\cap S(b_m)\cap S(b_n)\}\cup \cr
  & \qquad \cup \{S(b_i)\cap S(b_j)\cap
  S(b_n)-S(b_i)\cap S(b_j)\cap S(b_m)\cap S(b_n)\} \rbrack \cr
  &=S^{-1}\circ\lbrack \{S(b_i)\cap S(b_j)\cap S(b_m)-\bigl( S(b_i)\cup
S(b_j)\cup
  S(b_m)\bigr) \cap \bigl( S(b_i) \cap S(b_j)\cap S(b_n)\bigr) \} \cup \cr
  & \qquad \cup \{S(b_i)\cap S(b_j)\cap
  S(b_n)-\bigl( S(b_i)\cap S(b_j)\cap S(b_m)\bigr) \cap \bigl( S(b_i)\cap
S(b_j)\cap
  S(b_n)\bigr) \} \rbrack\cr
  &=S^{-1}\circ\lbrack \bigl( S(b_i)\cap S(b_j)\cap S(b_m)\bigr) \bigtriangleup
  \bigl( S(b_i)\cap S(b_j)\cap S(b_n)\bigr) \rbrack \cr
  &=\bigl( S^{-1}\circ \lbrack S(b_i)\cap S(b_j)\cap S(b_m)\rbrack \bigr)
\oplus
  \bigl( S^{-1}\circ \lbrack S(b_i)\cap S(b_j)\cap S(b_n)\rbrack \bigr). \cr}$$
Since by assumption, the  spin structure vectors in the big parenthesis in the
last
line of the above expression have an even number of entries 1, the spin
structure vector
which is their $mod\, 2$ sum also has an even number of entries 1.

{\bf Axiom $A_6$}: What we have to show is that if the triplets $(\chi_i, y_i,
\omega
_i)$, $i=1,\cdots ,6$ in the vectors $b_m$ and $b_n$  have the form required by
axiom
$A_6$ for any $m, n= 1,\cdots ,k$, then the same is true for the triplets in
the vector
$b_m\oplus b_n$. Let us list explicitly the allowed set of the triplets: for
the
Neveu-Schwarz sector the set $A=\{(0, 1, 1),\, (1, 0, 1),\,(1, 1, 0),\, (0, 0,
0)\}$ and
for the Ramond sector the set $B=\{(1, 0, 0),\, (0, 1, 0),\, (0, 0, 1),\, (1,
1, 1)\}$.
Let us notice that the spin structure vector $b_m\oplus b_n$ belongs to the
Neveu-Schwarz sector if and only if either $b_m$ and $b_n$ belong to the
Neveu-Schwarz
sector or $b_m$ and $b_n$ belong to the Ramond sector while  $b_m\oplus b_n$
belongs to
the Ramond sector if and only if either $b_m$ belongs to the Neveu-Schwarz
sector and
$b_n$ belongs to the Ramond sector or $b_m$ belongs to the Ramond sector and
$b_n$
belongs to the Neveu-Schwarz sector. That is why in order to complete the proof
it is
enough to check that:
  \item{i.}$ a\oplus a'\in A,\qquad \forall \, a, a'\in A;$
  \item{ii.}$b\oplus b'\in A, \qquad \forall \, b, b'\in B;$
  \item{iii.}$ a\oplus b\in B, \qquad \forall \, a\in A,\, \forall \, b\in B.$

\noindent It is easy to verify by direct computation that this is indeed the
case.

Now we turn to the Theorem, proving it also axiom by axiom. Before doing that,
we need
to introduce the following notations. Let: $ C(b)=\{ l:\, b^l\in  \rbrack -1,\,
0\lbrack
\, \cup \, \rbrack 0,\, 1\lbrack ,\enskip b\in \Xi \} $. Since all elements in
the set
$C(b)$ are indices of entries of a spin structure vector $b$ which correspond
to complex
fermions, for any $l'\in C(b)$ there is an $l''\in C(b)$ such that $ l'\ne l''$
and
$b^{l'}= b^{l''}$. We denote as $C'(b)$ a subset of $C(b)$ that contains only
one
element  $l'$ from each such pair $l'$ and $l''$.

{\bf Proof of the Theorem}: Let us notice first that the set of spin structure
vectors
$B'=\Lambda _{m,n}B$ for any $m,n= 1,\cdots ,k+1$ such that $m\ne k+1$, $n\ne
1$ and
$n\ne m$ is a basis, i.e., generates the group $\Xi$.

Let $N'_{k+1}$ stand for the least positive integer such that
$N'_{k+1}b_{k+1}'=0\,
mod\, 2$ where
$b'_{k+1}=b_{k+1}\oplus b_m$. Since $N_{k+1}$ is assumed to be even, it is easy
to see that $N'_{k+1}=N_{k+1}$.

{\bf Axiom $A_1$}: According to the Lemma for the case of axiom $A_1$, it
is enough to show that if the basis $B=\{b_i,\, i= 1,\cdots ,k+1\}$
satisfies axiom $A_1$ then the basis $B'=\Lambda_{m,k+1}B$ also does for any
$m= 1,\cdots ,k$. In order to do it, let us consider the following ``linear"
combination:
  $$a_1b_1\oplus a_2b_2\oplus\cdots\oplus a_{k+1}b'_{k+1}$$
where $a_i= 0,1$ for any $i= 1,\cdots ,k$ and $a_{k+1}= 0,\cdots , N_{k+1} -1$.
It is clear that if all $a_i=0$ then the above ``linear" combination is also
equal to
zero. Let us prove the converse statement. Let $\bar a_m = 0,1$ be such
that $\bar a_m=a_{k+1}\, mod\, 2$. It is easy to see that
$a_{k+1}b_m\mathop{=}\limits^{mod\, 2}\,\bar a_m b_m$. Taking into account this
fact, we
rewrite the above ``linear" combination as follows:
  $$a_1b_1\oplus\cdots\oplus(a_m\oplus \bar a_m)b_m\oplus\cdots
  \oplus a_{k+1}b_{k+1}.$$
Since by assumption the basis $B$ satisfies axiom $A_1$, from the equality of
the above
``linear" combination to zero it follows that $a_i=a_m\oplus\bar a_m=0$ for any
$i= 1,\cdots ,k+1$ such that $i\ne m$. Thus only $a_m$ is left to be shown to
be
equal to zero. But, it directly follows from the equality $a_m\oplus\bar a_m=0$
and from the fact that $\bar a_m=0$ since $a_{k+1}=0$.

{\bf Axiom 2}: Axiom $A_2$ obviously holds since the transformations $\Lambda
_{m,n}$
do not change the spin structure vector  $b_1$ due to the condition $n\ne 1$.

{\bf Axiom 3}: Since $N'_{k+1}=N_{k+1}$ and $N_{k+1}$ is even by assumption,
the least
common multiple of $N_i$ and $N'_{k+1}$ is equal to $N_{k+1}$ for any $i=
1,\cdots ,k$. Therefore, according to the Lemma for the case of axiom $A_3$,
what we
have to prove is that:
  $$\eqalignno{
  N_{k+1}(b_i; b_{k+1}\oplus b_m)&=0\, mod\, 4,\qquad \forall \, i, m=
  1,\cdots ,k&\cr
  N_{k+1}(b_m\oplus b_n; b_{k+1})&=0\, mod\, 4,\qquad \forall \, m, n= 1,\cdots
  ,k.\cr }$$
In order to do this it is enough to show that:
  $$N_{k+1}(b_i; b_{k+1}\oplus b_m)\mathop{=}\limits^{mod\, 4}\,
  N_{k+1}(b_i;b_{k+1}+b_m),\qquad \forall \, i, m= 1,\cdots ,k\eqno {(6)}$$
  $$N_{k+1}(b_m\oplus b_n; b_{k+1})\mathop{=}\limits^{mod\, 4}\,N_{k+1}(b_m+
  b_n; b_{k+1}),\qquad \forall \, m, n= 1,\cdots ,k.\eqno{(7)}$$
Indeed, due to the linearity of the inner product defined in (1) with
respect to the regular summation, we rewrite the right-hand sides of
equalities (6) and (7) as follows:
  $$\eqalignno{
  N_{k+1}(b_i; b_{k+1}+ b_m)&=N_{k+1}(b_i; b_{k+1})+N_{k+1}(b_i; b_m)&\cr
  N_{k+1}(b_m+ b_n;b_{k+1})&=N_{k+1}(b_m; b_{k+1})+N_{k+1}(b_n; b_{k+1}).\cr
}$$
The right-hand sides of the above expressions are equal to $0\, mod\, 4$ for
any
$i,m,n = 1,\cdots ,k$. Indeed, the first summand in the right-hand side of the
upper
expression and both summands in the right-hand side of the lower expression are
equal to
$0\, mod\, 4$ for any $i,m,n= 1,\cdots ,k$ because the vectors $b_i,b_m,b_n$
and
$b_{k+1}$ belong to the basis $B$, which is assumed to satisfy axiom $A_3$. The
second
summand in the right-hand side of the upper expression is equal to $0\, mod\,
4$
because it can be rewritten as $({N_{k+1}\over 2})2(b_i; b_m)$, where
$2(b_i;b_m)=0\,
mod\, 4$ for any $i,m= 1,\cdots ,k$ due to axioms $A_3$ and $A_4$ and
since $N_{k+1}\over 2$ is an integer.

Let us prove equality (6). In order to do it, we split the sum in the inner
product as
follows:
  $$\eqalignno{
  N_{k+1}(b_i; b_{k+1}\oplus b_m)&={N_{k+1}\over 2}\sum_l b_i^l(b_{k+1}^l\oplus
  b_m^l)&\cr
  &={ N_{k+1}\over 2}\mathop{\mathop{\sum}\limits
  _{l:\, b_m^l=1}}\limits_{b_{k+1}^l>0}\, b_i^l(b_{k+1}^l\oplus
b_m^l)+{N_{k+1}\over 2}
  \mathop{\mathop{\mathop{\mathop{\sum}\limits_{l:\,
b_m^l=1}}\limits_{b_{k+1}^l\le
  0}}\limits_{{\rm or:}\, b_m^l=0}}\limits_{{\rm any}\, b_{k+1}^l}\,
  b_i^l(b_{k+1}^l\oplus b_m^l)&\cr
  &={N_{k+1}\over 2}\mathop{\mathop{\sum}\limits
  _{l:\, b_m^l=1}}\limits_{b_{k+1}^l>0}\,b_i^l(b_{k+1}^l-b_m^l)+{N_{k+1}\over
2}
  \mathop{\mathop{\mathop{\mathop{\sum}\limits_{l:\,
b_m^l=1}}\limits_{b_{k+1}^l\le
  0}}\limits_{{\rm or:}\, b_m^l=0}}\limits_{{\rm any}\, b_{k+1}^l}
  b_i^l(b_{k+1}^l+b_m^l),\cr } $$
where in order to obtain the last equality, we used the obvious properties of
$mod\, 2$ summation $\oplus$. It is easy to see that the last line of  the
above
relation differs from the right-hand side of equality (6) by the following
expression:
 $$N_{k+1}\mathop{\sum}\limits_{l:\, b_m^l=b_i^l=1}\,b_i^l
 b_m^l,$$
which is equal to $0\, mod\, 4$ for any $i,m= 1,\cdots ,k$ as a product of an
even
$N_{k+1}$ and the sum. This sum is also even since each summand is equal to 1
and the
number of summands is even because it is equal to the number of common entries
1 in the
vectors $b_i$ and $b_m$ which is even for any $i,m= 1,\cdots ,k$ due to axiom
$A_5$.

Now let us prove equality (7). In order to do it we again split the sum in the
inner product as follows:
  $$\eqalignno{
  N_{k+1}(b_m\oplus b_n; b_{k+1})&={N_{k+1}\over 2}\sum_l(b_m^l\oplus
  b_n^l)b_{k+1}^l\cr
  &={N_{k+1}\over 2}\mathop {\sum}\limits_{l:\, b_m^l=b_n^l}\, (b_m^l\oplus
  b_n^l)b_{k+1}^l+{N_{k+1}\over 2}\mathop {\sum}\limits_{l:\, b_m^l\ne b_n^l}\,
  (b_m^l\oplus b_n^l)b_{k+1}^l.\cr}$$
We consider the sums in the last equality of the above expression separately.
The first sum is equal to zero, since the condition $b_m^l=b_n^l$ means that
either
$b_m^l=b_n^l=0$ or $b_m^l=b_n^l=1$, and in both cases $b_m^l\oplus b_n^l=0$.
For the
second sum it is easy to see that $b_m^l\oplus b_n^l=b_m^l+b_n^l$, since the
condition
$b_m^l\ne b_n^l$ means that either $b_m^l=1$ and $b_n^l=0$ or $b_m^l=0$
and $b_n^l=1$. Therefore, we obtain the following equality:
 $$N_{k+1}(b_m\oplus b_n, b_{k+1})={N_{k+1}\over 2}\mathop
 {\sum}\limits_{l:\, b_m^l\ne b_n^l}\, (b_m^l+b_n^l)b_{k+1}^l.$$
In order to finish the proof, it is enough to notice that the right-hand side
of the
above relation differs from the right-hand side of equality (7) by the
following
expression:
 $$N_{k+1}\mathop{\sum}\limits_{l:\, b_m^l= b_n^l=1} b_{k+1}^l=N_{k+1}
 \mathop{\mathop{\sum}\limits_{l:\,
b_m^l=b_n^l=1}}\limits_{b_{k+1}^l=1}\,b_{k+1}^l+
 N_{k+1}\mathop{\mathop {\sum}\limits_{l:\, b_m^l=b_n^l=1}}
 \limits_{b_{k+1}^l\ne 0,\, 1}\,b_{k+1}^l,$$
which is equal to $0\, mod\, 4$ for any $m,n = 1,\cdots ,k$. In order to show
this, we
consider the two summands in the right-hand side of the above expression
separately. The
first summand is equal to $0\, mod\, 4$ for any $m,n = 1,\cdots ,k$ as a
product of an
even $N_{k+1}$ and the indicated sum. This sum is also even since each summand
is equal
to 1 and the number of terms in the sum is even because it is equal to the
number of
common entries 1 in the vectors $b_m, b_n$ and $b_{k+1}$, which is even for any
$m,n=
1,\cdots ,k$ due to axiom $A_5$. The second summand is equal to $0\, mod\, 4$
for any
$m,n = 1,\cdots ,k$, since the condition $b_{k+1}^l\ne 0$ and $b_{k+1}^l\ne 1$
means
that we sum over the $l$ corresponding to complex fermions, which entries come
in
pairs, and hence this summand can be rewritten as:
 $$2\mathop{\mathop{\sum}\limits_{l:\, b_m^l=b_n^l=1}}
 \limits_{l\in C'(b_{k+1})}\, N_{k+1}b_{k+1}^l,$$
where each term in this sum is even by definition of $N_{k+1}$.

{\bf Axiom $A_4$}: First let us show that:
 $$N_{k+1}(b_{k+1}\oplus b_m+{\bf 1})\ne 0\, mod\, 4,\qquad \forall \, m=
1,\cdots
,k,$$
so the condition $N_{k+1}(b_{k+1}\oplus b_m; b_{k+1}\oplus b_m)=N_{k+1}({\bf
1};{\bf 1})\, mod\, 16$ is never applied. Let us assume the contrary, i.e.,
that there
is such $m= 1,\cdots ,k$ that:
 $$N_{k+1}(b_{k+1}\oplus b_m+{\bf 1})=0\, mod\, 4.\eqno{(8)}$$

We consider two following cases:

\item{1.}$N_{k+1}\ne 0\, mod\, 4$: We are going to show that in this case
equality (8)
contradicts axiom $A_1$. Indeed, since $N_{k+1}\over 2$ is odd, we obtain
${N_{k+1}\over 2}{\bf 1}\mathop{=}\limits^{mod\, 2}{\bf 1}$ and ${N_{k+1}\over
2}b_m
\mathop{=}\limits^{mod\, 2}b_m$. Therefore, since $b_1={\bf 1}$, equality (8)
implies
that the following ``linear" combination with non-zero coefficients
  $${N_{k+1}\over 2}b_{k+1}\oplus b_m \oplus b_1=0\, mod\, 2.$$

\item{2.}$N_{k+1}=0\, mod\, 4$: In this case equality (8) implies that
  $N_{k+1}b_{k+1}=0\, mod\, 4$. But, this relation contradicts the definition
of
  $N_{k+1}$ since, the integer $N_{k+1}\over 2$, which is smaller than
$N_{k+1}$,
satisfies the following equality:
  $${N_{k+1}\over 2}b_{k+1}=0\, mod\, 2.$$

In order to complete the proof, according to the Lemma for the case of axiom
$A_4$ and
since $N_{k+1}$ is assumed to be even, we have to show that:
  $$N_{k+1}(b_{k+1}\oplus b_m; b_{k+1}\oplus b_m)=0\, mod\, 8,\quad \forall \,
m=
  1,\cdots ,k.$$
In order to do this, it is enough to show that:
  $$N_{k+1}(b_{k+1}\oplus b_m; b_{k+1}\oplus b_m)\mathop {=}\limits^{mod\, 8}\,
  N_{k+1}(b_{k+1}+b_m; b_{k+1}+b_m),\quad \forall \, m= 1,\cdots ,k.
\eqno{(9)}$$
Indeed, due to the linearity of the inner product defined in (1) with respect
to
the regular summation, we rewrite the right-hand side of the above equality
as follows:
  $$N_{k+1}(b_{k+1}+b_m;
b_{k+1}+b_m)=N_{k+1}(b_{k+1};b_{k+1})+2N_{k+1}(b_{k+1};
  b_m)+N_{k+1}(b_m; b_m).$$
The right-hand side of the above expression is equal to $0\, mod\, 8$ for any
$m=
1,\cdots ,k$. Indeed the first summand is equal to $0\, mod\, 8$ since the
vector
$b_{k+1}$ belongs to the basis $B$, which is assumed to satisfy axiom $A_4$.
The third
summand is equal to $0\, mod\, 8$ for any $m= 1,\cdots ,k$, since it can be
rewritten
as $({N_{k+1}\over 2})2(b_m; b_m)$, where  $2(b_m; b_m)=0\, mod\, 8$ because
vector
$b_m$ belongs to the basis $B$, which is assumed to satisfy axiom $A_4$, and
$N_{k+1}\over 2$ is an integer. The second summand is equal to $0\, mod\, 8$
for any
$m= 1,\cdots ,k$ because $N_{k+1}(b_{k+1}; b_m)=0\, mod\, 4$ due to axiom
$A_3$.

Let us prove equality (9). In order to do it, we split the sum in the inner
product as
follows:
 $$\eqalignno{
  N_{k+1}&(b_{k+1}\oplus b_m; b_{k+1}\oplus b_m)=
  {N_{k+1}\over 2}\sum_l (b_{k+1}^l\oplus b_m^l)(b_{k+1}^l\oplus b_m^l)&\cr
  &={N_{k+1}\over 2}\mathop{\mathop{\sum}\limits_{l:\,
b_m^l=1}}\limits_{b_{k+1}^l>0}\,
  (b_{k+1}^l\oplus b_m^l)(b_{k+1}^l\oplus b_m^l)+{N_{k+1}\over 2}
  \mathop{\mathop{\mathop
  {\mathop{\sum}\limits_{l:\, b_m^l=1}}\limits_{b_{k+1}^l\le 0}}\limits_{{\rm
or:}\,
  b_m^l=0}}\limits_{{\rm any}\, b_{k+1}}\, (b_{k+1}^l\oplus
b_m^l)(b_{k+1}^l\oplus
  b_m^l)&\cr
  &={N_{k+1}\over 2}\mathop{\mathop{\sum}\limits_{l:\,
b_m^l=1}}\limits_{b_{k+1}^l>0}\,
  (b_{k+1}^l- b_m^l)(b_{k+1}^l-b_m^l)+{N_{k+1}\over 2} \mathop{\mathop{\mathop
  {\mathop{\sum}\limits_{l:\, b_m^l=1}}\limits_{b_{k+1}^l\le 0}}\limits_{{\rm
or:}\,
  b_m^l=0}}\limits_{{\rm any }\, b_{k+1}}\, (b_{k+1}^l+ b_m^l)(b_{k+1}^l+
b_m^l),\cr}$$
where, in order to obtain the last equality, we used the obvious properties of
$mod\, 2$ summation $\oplus$. In order to finish the proof, it is enough to
notice that
last line of the above relation differs from the right-hand side of equality
(9) by the
following expression:
  $$2N_{k+1}\mathop{\mathop{\sum}\limits_{l:\, b_m^l=1}}\limits_{b_{k+1}^l>0}\,
  b_{k+1}^lb_m^l= 2N_{k+1}\mathop{\mathop{\sum}\limits_{l:\, b_m^l=1}}
  \limits_{b_{k+1}^l=1}\,b_{k+1}^lb_m^l+2N_{k+1}\mathop{\mathop{\mathop{\sum}
  \limits_{l:\, b_m^l=1}}\limits_{b_{k+1}^l>0}}\limits_{b_{k+1}^l\ne
 1}\,b_{k+1}^lb_m^l,$$
which is equal to $0\, mod\, 8$ for any $m= 1,\cdots ,k$. In order to show
this, we
consider the two summands in the right-hand side of the above expression
separately. The
first summand is equal to $0\, mod\, 8$ for any $m= 1,\cdots ,k$ as a product
of 2, an
even $N_{k+1}$, and the indicated sum. This sum is also even, since each term
in the
sum is equal to 1 and the number of the terms is even because it is equal to
the number
of common entries 1 in the vectors $b_{k+1}$ and $b_m$, which is even for any
$m= 1,\cdots ,k$  due to axiom $A_5$. The second summand is equal to $0\, mod\,
8$
for any $m= 1,\cdots ,k$, since the condition $0<b_{k+1}^l<1$ means that we sum
over
the $l$ corresponding to complex fermions, which entries come in pairs, and
hence this
summand can be rewritten as:
  $$4\mathop{\mathop{\sum}\limits_{l:\, b_m^l=1}}\limits_{l\in  C'(b_{k+1})}
  \, N_{k+1}b_{k+1},$$
 where each term in the sum is even by definition of $N_{k+1}$.

{\bf Axiom $A_5$}: Since this axiom is a condition only for entries of spin
structure
vectors equal to 1, the proof is completely analogous to the proof of axiom
$A_5$ in the
Lemma.

{\bf Axiom $A_6$}: Since this axiom is a condition for the entries $l= 3,\cdots
,20$, where the non-integer vector $b_{k+1}$ has the same structure as integer
vectors
$b_i,\, i= 1,\cdots ,k$, the proof is completely analogous to the proof of
axiom
$A_6$ in the Lemma.

\bigskip
\centerline{\bf Conclusion}

In the presented paper we consider a basis $B=\{b_i,\, i= 1,\cdots ,k+1\}$
generating a finite abelian spin structure group $\Xi$ in four-dimensional
superstring
theory in free fermionic formulation. According to the most successful
realistic
models, the basis spin structure vectors $b_i,\, i= 1,\cdots ,k$ are assumed to
have
only entries 0 or 1, while  $b_{k+1}$ is allowed to have any rational entries
belonging
to $]-1,1]$. We prove that axioms $A_1-A_6$ for the basis B are invariant with
respect
to the transformation $\Lambda _{m,n}$ that replaces a basis spin structure
vector
$b_n$ by a vector which is a $mod\, 2$ sum of vectors $b_m$ and $b_n$ for any
$m\ne k+1$, $n\ne1$ and $m\ne n$, and leaves the rest of the basis spin
structure
vectors unchanged. Since the transformations $\Lambda _{m,n}$ for different
$m,n$
generate all non degenerate transformations of the basis $B$ that preserve the
spin
structure vector $b_1$ and a single spin structure vector $b_{k+1}$ with
general
rational entries, we conclude that these axioms are conditions for the whole
group
$\Xi$ and not just conditions for a particular choice of its basis. Hence,
these
transformations generate the discrete symmetry group of four-dimensional
superstring
models in free fermionic formulation. This is the physical meaning of the
obtained
result. The practical impact is that in searching for a realistic model in
four-dimensional superstring theory it is enough to search only through
different
groups $\Xi$ which number dramatically less than the number of their different
bases.

In addition we would like also to point out that axiom $A_1$ is the only axiom
that is
not satisfied in the generalization of the Theorem for the case of a basis $B$
with
arbitrary number of non-integer spin structure vectors.

\bigskip
\centerline {\bf Acknowledgments }

I would like to thank Professor John S. Hagelin and Dr.\ Stephen Kelley for
stimulating
discussions. Also I would like to thank my wife Svetlana for typing the
manuscript and
Dr. Geoffrey Golner and Sherri Shields for editing the manuscript.

\references

\rfrnc I. Antoniadis, C. Bachas and C. Kounnas, {\it Nuclear Physics}, {\bf
289B}
(1987) 87.//

\rfrnc I. Antoniadis and C. Bachas, {\it Nuclear Physics}, {\bf 298B}
(1988) 586.//

\rfrnc H. Dreiner, J. L. Lopez, D.V. Nanopoulos and D. Reiss, {\it Nuclear
Physics},
{\bf B320} (1989) 401.//

\rfrnc H. Kawai, D.C. Lewellen, J.A. Schwarz and S.-H.H. Tye, {\it Nuclear
Physics},
{\bf B320} (1989) 431.//

\rfrnc I. Antoniadis, J. Ellis, J.S. Hagelin and D.V. Nanopoulos, {\it Physics
Letters}, {\bf B231} (1989) 65.//

\rfrnc J.L. Lopez, D.V. Nanopoulos and K. Yuan, {\it Nuclear Physics}, {\bf
B399} (1993)
654.//

\rfrnc S. Kalara, J.L. Lopez and D.V. Nanopoulos, {\it Physics Letters}, {\bf
B245}
(1990) 421.//

\rfrnc I. Antoniadis, G.K. Leontaris and J. Rizos, {\it Physics Letters}, {\bf
B245}
(1990) 161.//

\end